\documentstyle[manuscript,aps]{revtex}
\begin{document}
\title{Kantowski-Sachs String Cosmologies }
\author{John D. Barrow \footnote{E-mail:J.D.Barrow@sussex.ac.uk}}
\address{{\it Astronomy Centre, University of Sussex, Brighton BN1 9QH, U.K.,}\\
Mariusz P. D\c{a}browski \footnote{E-mail:mpd@wmf.univ.szczecin.pl}\\
{\it Astronomy Centre, University of Sussex, Brighton BN1 9QH, U.K.,}\\
Institute of Physics, University of Szczecin, Wielkopolska 15, 70-451
Szczecin, Poland}
\date{\today}
\maketitle

\begin{abstract}
We present new exact solutions of the low-energy-effective-action string
equations with both dilaton $\phi $ and axion $H$ fields non-zero. The
background universe is of Kantowski-Sachs type. We consider the possibility
of a pseudoscalar axion field $h$ ($H=e^\phi (dh)^{*}$) that can be either
time or space dependent. The case of time-dependent $h$ reduces to that of a
stiff perfect-fluid cosmology. For space-dependent $h$ there is just one
non-zero time-space-space component of the axion field $H$, and this
corresponds to a distinguished direction in space which prevents the models
from isotropising. Also, in the latter case, both the axion field $H$ and
its tensor potential $B$ ($H=dB$) are dependent on time and space yet the
energy-momentum tensor remains time-dependent as required by the homogeneity
of the cosmological model. 
\end{abstract}

\pacs{PACS numbers: 98.80.Cq \hspace*{1mm} 
,  hep-th/9608136}

\preprint{SUSSEX-AST 96/...., hep-th/9608136}

\renewcommand{\theequation}{\thesection.\arabic{equation}}

\newpage

\section{Introduction}

\setcounter{equation}{0}

The motion of the bosonic string in background fields is governed by the
action for the nonlinear sigma model \cite{frad,call}. Mueller \cite{mel}
has solved both zeroth and first-order inverse string tension $\alpha
^{\prime }$ equations for multi-dimensional Bianchi I cosmological model
without the antisymmetric axion field. The isotropic cosmological
backgrounds without axion have been extensively studied by Gasperini and
Veneziano \cite{ven,gave,gave1,gave2}, and special attention has been paid
to the pre-Big-Bang solutions in relation to scale-factor duality. The
homogeneous axion-dilaton cosmology gives rise to the question \cite{ed} of
whether the antisymmetric 3-index axion field strength $H_{\alpha \beta
\gamma },$ ($\alpha ,\beta ,\gamma =0,1,2,3),$ or its antisymmetric 2-index
tensor potential $B_{\beta \gamma }$ ($H_{\alpha \beta \gamma }=6\partial
_{[\alpha }B_{\beta \gamma ]}$) should be homogeneous. By analogy with the
Einstein-Maxwell equations, most investigators have considered an
homogeneous (time-dependent) axion field strength $H$ \cite{ed1,ed,bk,gasp},
but Copeland et al. \cite{ed} assumed that it was the potential $B$ which
should be homogeneous. Actually, because of the antisymmetry of the axion
field, one has either of the two possibilities: the components $H_{oij},$ $%
(i,j=1,2,3),$ vanish if the potential $B_{ij}$ is space-dependent and the
components $H_{ijk}$ vanish if the potential $B_{ij}$ is time-dependent.
Copeland et al. \cite{ed} framed the problem in terms of the dual
pseudoscalar axion field $h$ ($H=e^\phi (dh)^{*}$, where $*$ is the
spacetime dual) which was taken to be time or space dependent,
respectively). They  concluded that for time-dependent antisymmetric tensor
potential $B_{ij}$ there exists just one non-zero component of the axion
field, $H_{0ij},i,j=1,2,3,$ and this gives rise to Bianchi I universes which
cannot isotropise at late times. Similarly, Barrow and Kunze \cite{bar1}
have classified the degrees of freedom available to the antisymmetric field
strength $H$ in all Bianchi type spacetimes, assuming time-dependence of the
tensor potential $B$ as well in the orthonormal-frame formalism. Other
papers about the evolution of the axion field have also appeared \cite
{saar,gal}.

In this paper we will consider a spatially-homogeneous background spacetime
of Kantowski-Sachs type. This is the only spatially homogeneous universe
that is not included in the Bianchi classification. It falls outside this
classification of models with three-dimensional isometry groups because it
possesses a four-dimensional group of motions with no simply-transitive
three-dimensional subgroup. We consider both time-dependent and
space-dependent pseudoscalar axion field $h$ (cf. Appendix). In the former
case we have effectively another scalar field (or equivalently, a stiff
perfect-fluid) cosmology. In the latter case we produce a model which
evolves like that given by Copeland et al. \cite{ed} for Bianchi I models.

The field equations will describe three different anisotropic 3+1
dimensional spacetimes. Those with zero and negative curvature are just
axisymmetric Bianchi type I and III universes. The positive curvature models
constitute the Kantowski-Sachs models (first found by Kompanyeets and
Chernov \cite{kom}). They are closed anisotropic universes. In the special
case where they become isotropic they reduce to the closed
Friedmann-Robertson-Walker universes.

We give solutions for models with all curvatures. For a time-dependent
pseudoscalar axion field $h$ we give a parametric solution of the
low-energy-effective-action equations for the system containing both dilaton
and axion in Section II. We also give an explicit special solution in terms 
of the cosmic time when the axion field is absent in Section III. Following 
the discussion of Copeland et al. \cite{ed} in Section IV, we examine whether 
it is possible to employ a
time-independent pseudoscalar axion field $h$ in Kantowski-Sachs geometries
and, if so, which of its components are allowed to be nonzero. In the Appendix 
we also discuss some relations between our work and these earlier studies.

\section{Low-energy-effective-action equations and solutions with
time-dependent pseudoscalar axion field}

\setcounter{equation}{0}

The low-energy-effective-action field equations are given by \cite{gasp} 
\begin{eqnarray}
R_\mu ^\nu +\nabla _\mu \nabla ^\nu \phi -\frac 14H_{\mu \alpha \beta
}H^{\nu \alpha \beta } &=&0, \\
R-\nabla _\mu \phi \nabla ^\mu \phi +2\nabla _\mu \nabla ^\mu \phi -\frac 1{%
12}H_{\mu \nu \beta }H^{\mu \nu \beta } &=&0, \\
\partial _\mu \left( e^{-\phi }\sqrt{-g}H^{\mu \nu \alpha }\right)  &=&0,
\end{eqnarray}
where $\phi $ is the dilaton field, $H_{\mu \nu \beta }=6\partial _{[\mu
}B_{\nu \beta ]}$ is the strength of the antisymmetric tensor field $B_{\mu
\nu }=-B_{\nu \mu }$ is its antisymmetric tensor potential. We choose the
metric of spacetime to be of Kantowski-Sachs form, with \cite{KS} 
\begin{equation}
ds^2=dt^2-X^2(t)dr^2-Y^2(t)d\Omega _k^2,
\end{equation}
where the angular metric is 
\begin{eqnarray}
d\Omega _k^2 &=&d\theta ^2+S^2(\theta )d\psi ^2,  \nonumber \\
S(\theta ) &=&\left\{ 
\begin{array}{l}
\sin {\theta }\hspace{0.5cm}{\rm for}\hspace{0.3cm}k=+1, \\ 
\theta \hspace{0.5cm}{\rm for}\hspace{0.3cm}k=0, \\ 
\sinh {\theta }\hspace{0.5cm}{\rm for}\hspace{0.3cm}k=-1,
\end{array}
\right. \ 
\end{eqnarray}
and $X$ and $Y$ are the expansion scale factors. We shall consider models of
all three curvatures in the same analysis. Strictly, only the $k=+1$ models
fall outside the Bianchi classification but we shall refer to them all as
Kantowski-Sachs models for simplicity. The nonzero Ricci tensor components
are 
\begin{eqnarray}
-R_0^0 &=&\frac{\ddot X}X+2\frac{\ddot Y}Y, \\
-R_1^1 &=&\frac{\ddot X}X+2\frac{\dot X}X\frac{\dot Y}Y, \\
-R_2^2 &=&-R_3^3=\frac{k+\dot Y^2}{Y^2}+\frac{\ddot Y}Y+\frac{\dot X}X\frac{%
\dot Y}Y,
\end{eqnarray}
and the scalar curvature is 
\begin{equation}
-R=2\frac{\ddot X}X+4\frac{\ddot Y}Y+2\frac{k+\dot Y^2}{Y^2}+4\frac{\dot X}X%
\frac{\dot Y}Y.
\end{equation}
Since the metric is spatially homogeneous the dilaton field can only depend
on time and we have 
\begin{equation}
\nabla _\mu \nabla ^\nu \phi =\phi _{,\mu }^{,\nu }+\Gamma _{\mu \rho }^\nu
\phi ^{,\rho },
\end{equation}
so 
\begin{eqnarray}
\nabla _0\nabla ^0\phi  &=&\ddot \phi , \\
\nabla _1\nabla ^1\phi  &=&\frac{\dot X}X\dot \phi , \\
\nabla _2\nabla ^2\phi  &=&\nabla _3\nabla ^3\phi =\frac{\dot Y}Y\dot \phi ,
\\
\nabla _0\phi \nabla ^0\phi  &=&\dot \phi ^2.
\end{eqnarray}
For the torsion field, $H_{\alpha \beta \gamma },$ we assume the simple
ansatz, similar to that of Batakis and Kehagias \cite{bk}, that $H_{\alpha
\beta \gamma }$ takes the form 
\begin{equation}
H_{123}=AS(\theta )
\end{equation}
where $A=$ constant and the rest of the components are taken to be zero.
This ansatz corresponds to a space-dependent antisymmetric potential $B_{\mu
\nu }=B_{\mu \nu }(r,\theta ,\psi )$ or to a time-dependent pseudoscalar
axion field $h$ (cf. Appendix A - Eq. (A.5)). It is interesting to note that
the antisymmetric tensor potential components for (II.15) are given by $%
B_{12}=A\psi \sin {\theta },B_{23}=Ar\sin {\theta },B_{31}=-A\cos {\theta }$
for $k=+1$; $B_{12}=A\psi \sinh {\theta },B_{23}=Ar\sinh {\theta }%
,B_{31}=A\cosh {\theta }$ for $k=-1$ and $B_{12}=A\psi
,B_{23}=Ar,B_{31}=A\theta $ for $k=0$. With the choice (II.15) the field
equations (II.1) become 
\begin{eqnarray}
\frac{\ddot X}X+2\frac{\ddot Y}Y-\ddot \phi  &=&0, \\
\frac{\ddot X}X+2\frac{\dot X}X\frac{\dot Y}Y-\dot \phi \frac{\dot X}X-\frac 
12\frac{A^2}{X^2Y^4} &=&0, \\
\frac{k+\dot Y^2}{Y^2}+\frac{\ddot Y}Y+\frac{\dot X}X\frac{\dot Y}Y-\dot \phi
\frac{\dot Y}Y-\frac 12\frac{A^2}{X^2Y^4} &=&0,
\end{eqnarray}
The sum of (II.16) and (II.17) added to twice (II.18) gives 
\begin{equation}
2\frac{\ddot X}X+4\frac{\ddot Y}Y+2\frac{k+\dot Y^2}{Y^2}+4\frac{\dot X}X%
\frac{\dot Y}Y-\ddot \phi -\left( \frac{\dot X}X+2\frac{\dot Y}Y\right) \dot 
\phi -\frac 32\frac{A^2}{X^2Y^4}=0.
\end{equation}
The field equation (II.2) reads 
\begin{equation}
-2\frac{\ddot X}X-4\frac{\ddot Y}Y-2\frac{k+\dot Y^2}{Y^2}-4\frac{\dot X}X%
\frac{\dot Y}Y+2\ddot \phi -\dot \phi ^2+2\left( \frac{\dot X}X+2\frac{\dot Y%
}Y\right) \dot \phi +\frac 12\frac{A^2}{X^2Y^4}=0,
\end{equation}
so from the sum of (II.19) and (II.20) we have 
\begin{equation}
\ddot \phi -\dot \phi ^2+\left( \frac{\dot X}X+2\frac{\dot Y}Y\right) \dot 
\phi -\frac{A^2}{X^2Y^4}=0.
\end{equation}
At this stage we introduce a new time coordinate 
\begin{equation}
dt=XY^2e^{-\phi }d\tau ,
\end{equation}
and then ({II.21) }becomes 
\begin{equation}
\phi _{,\tau \tau }-A^2e^{-2\phi }=0,
\end{equation}
which solves as 
\begin{equation}
e^\phi =\cosh {\alpha \tau }+\sqrt{1-\frac{A^2}{\alpha ^2}}\sinh {\alpha
\tau },
\end{equation}
with $\alpha $ constant $(\alpha ^2>A^2)$. If we turn off the $H_{\mu \nu
\rho }$ field, the solution of (II.24) is simply 
\begin{equation}
\phi (\tau )=\alpha \tau +\gamma ,
\end{equation}
with $\gamma \ $a constant, which we may set to zero without loss of
generality. A useful relation, implied by (II.24), is 
\begin{equation}
\phi _{,\tau \tau }+\phi _{,\tau }^2=\alpha ^2.
\end{equation}
Using the time coordinate (II.22), equations (II.16)-(II.18) become 
\begin{eqnarray}
\left( \frac{X_{,\tau }}X\right) _{,\tau }+2\left( \frac{Y_{,\tau }}Y\right)
_{,\tau }-2\frac{Y_{,\tau }}Y\left( \frac{Y_{,\tau }}Y+2\frac{X_{,\tau }}X%
\right) +2\phi _{,\tau }\left( \frac{X_{,\tau }}X+2\frac{Y_{,\tau }}Y\right)
-\phi _{,\tau \tau }-\phi _{,\tau }^2 &=&0, \\
\left( \frac{X_{,\tau }}X\right) _{,\tau }-\frac 12\phi _{,\tau \tau } &=&0,
\\
\left( \frac{Y_{,\tau }}Y\right) _{,\tau }-\frac 12\phi _{,\tau \tau
}+kX^2Y^2e^{-2\phi } &=&0.
\end{eqnarray}
The equations (II.29) and (II.30) can be rewritten as 
\begin{eqnarray}
\left( \ln {X^2e^{-\phi }}\right) _{,\tau \tau } &=&0, \\
\left( \ln {Y^2e^{-\phi }}\right) _{,\tau \tau }+2kX^2Y^2e^{-2\phi } &=&0.
\end{eqnarray}
From (II.20), we see that the constraint equation (II.16) can be rewritten
as 
\begin{equation}
\frac{k+\dot Y^2}{Y^2}+2\frac{\dot X}X\frac{\dot Y}Y+\frac 12\dot \phi ^2-%
\dot \phi \left( \frac{\dot X}X+2\frac{\dot Y}Y\right) -\frac 14\frac{A^2}{%
X^2Y^4}=0,
\end{equation}
so, in terms of the time coordinate (II.22), 
\begin{equation}
\frac 12\left( \ln {X^2e^{-\phi }}\right) _{,\tau }\left( \ln {Y^2e^{-\phi }}%
\right) _{,\tau }+\left( \ln {Y}\right) _{,\tau }\left( \ln {Y}-\phi \right)
_{,\tau }+kX^2Y^2e^{-2\phi }=\frac 14A^2e^{-2\phi }.
\end{equation}
The solution of (II.30) is 
\begin{equation}
X^2e^{-\phi }=X_0e^{p\tau },
\end{equation}
with $X_0$ and $p$ constants. Then, from (II.25) and (II.35), for $A\neq 0,$
we have 
\begin{equation}
X(\tau )=\sqrt{X_0}e^{\frac 12p\tau }\sqrt{\cosh {\alpha \tau }+\sqrt{1-%
\frac{A^2}{\alpha ^2}}\sinh {\alpha \tau }},
\end{equation}
or, for $A=0,$ 
\begin{equation}
X(\tau )=\sqrt{X_0}e^{\frac 12\left( p+\alpha \right) \tau }.
\end{equation}
The solution of (II.31) for the scale factor $Y$ is given by 
\begin{equation}
Y(\tau )=\frac 1{\sqrt{X_0}}e^{-\frac 12p\tau }\sqrt{\cosh {\alpha \tau }+%
\sqrt{1-\frac{A^2}{\alpha ^2}}\sinh {\alpha \tau }}\sqrt{M(\tau )},
\end{equation}
where $M(\tau )=X^2Y^2\exp {(-2\phi )}$ satisfies 
\begin{equation}
\left( \ln {M}\right) _{,\tau \tau }+2kM=0;
\end{equation}
hence, 
\begin{equation}
\frac 1{\sqrt{M(\tau )}}=\cosh {\beta \tau }+\sqrt{1-\frac k{\beta ^2}}\sinh 
{\beta \tau }   ,
\end{equation}
and $\beta^2 > k$. 
The constraint equation (II.33) may be now rewritten in terms of $M$ as 
\begin{equation}
\frac 14\left( \frac{M_{,\tau }}M\right) ^2+kM=\frac 14\left( \alpha
^2+p^2\right) ,
\end{equation}
which gives the condition 
\begin{equation}
\beta ^2=\frac 14\left( \alpha ^2+p^2\right) .
\end{equation}
Although $\tau$ is a parametric time related to the cosmic time by (II.22) we
give some plots of the scale factors $X$ and $Y$ (Eqs. II.35, II.37) and the
dilaton $\phi$ (Eq. II.24) in Figs.1-3. Different plots are given for different
values of the constants $\alpha$ and $p$, which reflects the string duality
symmetry here. As for $Y$ we give just the plot for curvature index $k = +1$.
Note that the value of the constant $\beta$ is constrained by Eq. (II.41) and
that $alpha$ and $p$ do not have to be positive. Also note that despite $X$ and
$\phi$ the plots of $Y(\tau)$ are time asymmetric in $\tau$. This refers to the
string duality which is more sophisticated for homogeneous models as it has
been shown for Bianchi I models in \cite{ed}.

\vspace{2.0cm}

\begin{center}
Figures 1-3
\end{center}

\vspace{2.0cm}

\section{Deparametrised solutions for time-dependent pseudoscalar axion field
}

\setcounter{equation}{0}

From the results of Section II we see that without the axion field $(A=0)$
the solutions for $\phi ,$ $X$ and $Y$ reduce to 
\begin{eqnarray}
\phi (\tau ) &=&\alpha \tau , \\
X(\tau ) &=&\sqrt{X_0}e^{\frac 12(\alpha +p)\tau }, \\
Y(\tau ) &=&\frac 1{\sqrt{X_0}}e^{\frac 12(\alpha -p)\tau }\tilde M,
\end{eqnarray}
where $\tilde M(\tau )$, the solution of (II.38), is given by 
\begin{equation}
\tilde M=\left\{ 
\begin{array}{l}
\beta \cosh ^{-1}\left( \beta \tau +\delta \right) \hspace{0.5cm}{\rm for}%
\hspace{0.3cm}k=+1, \\ 
\exp {(\beta \tau +\delta )}\hspace{1.0cm}{\rm for}\hspace{0.3cm}k=0, \\ 
\beta \sinh ^{-1}\left( \beta \tau +\delta \right) \hspace{0.5cm}{\rm for}%
\hspace{0.3cm}k=-1,
\end{array}
\right. \ 
\end{equation}
where $\delta \ $is constant, with the constraint given by (II.40).

From (II.22) and (III.1)-(III.3) we find that the time parameter in the
string frame is 
\begin{equation}
t(\tau )=\frac 1{\sqrt{X_0}}\int {e^{-\frac 12(\alpha -p)\tau }\tilde M%
^2d\tau }.
\end{equation}
Hence, this relation is integrable for $k\neq 0$, provided $\alpha =p$ (that
is, from (II.41), if $\beta ^2=\alpha ^2/2$). In this case we have 
\begin{eqnarray}
t(\tau )=\frac{\alpha ^2}{2\sqrt{X_0}}\left\{ 
\begin{array}{l}
\pm \frac{\sqrt{2}}\alpha \tanh {\left( \pm \frac \alpha {\sqrt{2}}\tau
+\delta \right) ,}\hspace{0.2cm}k=+1, \\ 
\mp \frac{\sqrt{2}}\alpha \coth {\left( \mp \frac \alpha {\sqrt{2}}\tau
+\delta \right) ,}\hspace{0.2cm}k=-1,
\end{array}
\right. \ 
\end{eqnarray}
For $k=0$ it is always integratable and gives 
\begin{equation}
t(\tau )=\frac 1{X_0s}\exp {(-s\tau -2\delta )},
\end{equation}
where 
\begin{equation}
s=-\frac 12\left( \alpha -p+4\beta \right) .
\end{equation}
After deparametrisation, equations (III.1)-(III.3) provide a simple solution
of (II.16)-(II.18) for $A=0$ and $\dot \phi =\dot X/X$. When $k\neq 0,$ it
is given by 
\begin{eqnarray}
X(t) &=&\left( k\frac{\frac \alpha {\sqrt{2}}-t}{\frac \alpha {\sqrt{2}}+t}%
\right) ^{\frac 1{\sqrt{2}}}, \\
Y(t) &=&\sqrt{k\left( \frac{\alpha ^2}2-t^2\right) }, \\
\phi (t) &=&\ln {\left( k\frac{\frac \alpha {\sqrt{2}}-t}{\frac \alpha {%
\sqrt{2}}+t}\right) ^{\frac 1{\sqrt{2}}}}
\end{eqnarray}
where the time coordinate has the ranges 
\begin{eqnarray}
0 &\leq &t\leq \frac{\alpha ^2}2\hspace{0.5cm}{\rm for}\hspace{0.3cm}k=+1, \\
t &\geq &\frac{\alpha ^2}2\hspace{0.5cm}{\rm for}\hspace{0.3cm}k=-1.
\end{eqnarray}
The volume expansion is given by 
\begin{equation}
V(t)=XY^2=\left[ k\left( \frac \alpha {\sqrt{2}}-t\right) \right] ^{\frac{%
\sqrt{2}+1}{\sqrt{2}}}\left( \frac \alpha {\sqrt{2}}+t\right) ^{\frac{\sqrt{2%
}-1}{\sqrt{2}}},
\end{equation}
and its evolution is qualitatively the same as that of the scale factor $Y(t)
$.

For $k=0$ we have 
\begin{eqnarray}
X(t) &=&\sqrt{X_0}\left( X_0st\right) ^{\frac{\alpha +p}{\alpha -p+4\beta }},
\\
Y(t) &=&\frac 1{X_0}\left( X_0st\right) ^{\frac{\alpha -p+2\beta }{\alpha
-p+4\beta }}, \\
\phi (t) &=&\frac{2\alpha }{\alpha -p+4\beta }\ln {(X_0st)}.
\end{eqnarray}
The plots of $X(t),Y(t)$ and $\phi (t)$ for $k=\pm 1$ are given in Figs.4-6. 
It is interesting to note that the scale factors and the dilaton remain the
same after the change $\alpha \rightarrow - \alpha$ and $t \rightarrow - t$
which refers to the string duality symmetry here \cite{ven}.
 
\vspace{2.0cm}

\begin{center}
Figures 4-6
\end{center}

\vspace{2.0cm}

For $k=+1,$ the universe starts at a cigar singularity with $X=\infty ,Y=0$
and terminates at a point singularity with $X=Y=0,$ \cite{col}. For $k=-1,$
the universe either starts at $t=\alpha ^2/2$ with a point singularity with
the ensuing volume expansion going to infinity (with asymptotic value of $X=1
$ for $t\rightarrow \infty $), or it starts with infinite volume (with $X$
taken to be equal to one at minus infinity) and collapses to a cigar
singularity.

In order to develop these solutions in the Einstein frame we need to change
the scale factors $X$ and $Y,$ together with the time coordinate, to 
\begin{eqnarray}
\tilde X &=&e^{-\frac \phi 2}X, \\
\tilde Y &=&e^{-\frac \phi 2}Y, \\
d\tilde t &=&e^{-\frac \phi 2}dt=\left( k\frac{\frac \alpha {\sqrt{2}}-t}{%
\frac \alpha {\sqrt{2}}+t}\right) ^{\frac 1{2\sqrt{2}}}dt.
\end{eqnarray}
The calculations for the $k=0$ (Bianchi I) case have already been given in 
\cite{ed}.

\section{Solutions with time-independent pseudoscalar axion field}

\setcounter{equation}{0}

In this section we consider the space-dependent pseudoscalar axion field $h$%
. This requirement, however, does not strictly correspond to the condition
that the antisymmetric tensor field strength $H_{\mu \nu \alpha }$ and its
antisymmetric potential $B_{\mu \nu }$ are space dependent as was the case
in the Bianchi I calculations of Copeland et al. \cite{ed} and the
Bianchi-type universes studied by Barrow and Kunze \cite{bar1}. If we have $%
B_{\mu \nu }=B_{\mu \nu }(t)$, then because of the antisymmetry 
\begin{equation}
H_{ijk}=0\text{ and }H_{oij}\neq 0.
\end{equation}
As we will show later, this is not the case in our formulation and it arises
from the fact that we do not use orthonormal frames. If we consider a
space-dependent pseudoscalar axion field, $h=h(r,\theta ,\psi )$, then
relation (A.1) of Appendix A requires $H^{123}=\epsilon ^{1230}e^\phi
\partial _0h=0$ and only the $H_{0ij}$ components of the axion field $H$ can
be non-zero (a similar situation  to that considered in refs. \cite{ed,bar1}%
).

There are some conditionswhich must be satisfied if the Kantowski-Sachs
equations (II.1)-(II.3) are to admit the axion field into the
low-energy-effective-action. One is that the off-diagonal components of $%
H_{\mu \lambda \sigma }H^{\nu \lambda \sigma }$ in Eq. (II.2) should vanish
and we have the condition 
\begin{equation}
g^{jj}g^{mm}H_{iom}H_{jom}=0\hspace{0.5cm}(i\neq j,{\rm no}\hspace{0.2cm}%
{\rm sum}),
\end{equation}
which means that only one of $H_{012},H_{023},$ or $H_{013}$ may be non-zero.

Suppose we choose $H_{012}\neq 0$. From the equation of motion (II.3) we
obtain 
\begin{equation}
H^{012}=\frac{Be^\phi }{XY^2S(\theta )},
\end{equation}
with $B\ $constant, and $S(\theta )$ given by (II.5), so 
\begin{equation}
H_{012}=-\frac{BXe^\phi }{S(\theta )}
\end{equation}
One can easily check that the integrability condition is fulfilled, since $%
\partial _3H_{012}=0$. However, (II.1) gives 
\begin{equation}
H^2\equiv H_{\mu \nu \lambda }H^{\mu \nu \lambda }=-6B^2\frac{e^{2\phi }Y^2}{%
S^2(\theta )}
\end{equation}
and there is explicit dependence on the spatial coordinate $\theta $, which
means that the ansatz $H_{012}\neq 0$ is inconsistent with the geometries
under consideration.

The next possibility is $H_{013}\neq 0$. This means that 
\begin{equation}
H^{013}=\frac{Ce^\phi }{XY^2S(\theta )}
\end{equation}
and 
\begin{equation}
H_{013}=Ce^\phi XS(\theta ),
\end{equation}
with $C$ constant. Then 
\begin{equation}
H^2=\frac{C^2e^{2\phi }}{Y^2},
\end{equation}
which does depend just on time. However, in this case we see that the
integrability condition $\partial _2H_{013}=0$ is not fulfilled and so the
choice $H_{013}\neq 0$ is also impossible.

The last possibility is given by 
\begin{equation}
H_{023}=De^\phi \frac{Y^2}XS(\theta )=Y^4S^2(\theta )H^{023},
\end{equation}
where $D$ is constant. This time-integrability condition $\partial
_1H_{023}=0$ is fulfilled and 
\begin{equation}
H^2=6D^2\frac{e^{2\phi }}{X^2},
\end{equation}
which depends only on the time coordinate, as required. Thus, $H_{023}\neq 0$
provides the only consistent choice. This naturally gives the space
dependence of the tensor potential $B$ because $H_{023}=\partial _0B_{23}$
(due to the gauge transformation we can eliminate all the components $B_{0i}$
\cite{ed}), so $B_{23}=DS(\theta )\int {Y^2X^{-1}e^\phi dt}$ and $%
B_{12}=B_{13}=0$. This also shows that $H_{123}$ component of the axion
field must vanish ($H_{312}=\partial _3B_{12}=0,H_{231}=\partial
_2B_{13}=0,H_{123}=\partial _1B_{23}(t,\theta )=0$. The final conclusion is
that our ansatz (IV.9) is correct.

Using this choice, the field equations (II.1-II.2) become 
\begin{eqnarray}
\frac{\ddot X}X+2\frac{\ddot Y}Y-\ddot \phi &=&-\frac{D^2e^{2\phi }}{2X^2},
\\
\frac{\ddot X}X+2\frac{\dot X}X\frac{\dot Y}Y-\dot \phi \frac{\dot X}X &=&0,
\\
\frac{k+\dot Y^2}{Y^2}+\frac{\ddot Y}Y+\frac{\dot X}X\frac{\dot Y}Y-\dot \phi
\frac{\dot Y}Y &=&-\frac{D^2e^{2\phi }}{2X^2},
\end{eqnarray}
The sum of (IV.9) and (IV.10) together with twice (IV.11) gives 
\begin{equation}
2\frac{\ddot X}X+4\frac{\ddot Y}Y+2\frac{k+\dot Y^2}{Y^2}+4\frac{\dot X}X%
\frac{\dot Y}Y-\ddot \phi -\left( \frac{\dot X}X+2\frac{\dot Y}Y\right) \dot 
\phi -\frac 32\frac{D^2e^{2\phi }}{X^2}=0.
\end{equation}
But equation (II.2) is 
\begin{equation}
-2\frac{\ddot X}X-4\frac{\ddot Y}Y-2\frac{k+\dot Y^2}{Y^2}-4\frac{\dot X}X%
\frac{\dot Y}Y+2\ddot \phi -\dot \phi ^2+2\left( \frac{\dot X}X+2\frac{\dot Y%
}Y\right) \dot \phi -\frac 12\frac{D^2e^{2\phi }}{X^2}=0,
\end{equation}
so from the sum of (IV.12) and (IV.13) we have 
\begin{equation}
\ddot \phi -\dot \phi ^2+\left( \frac{\dot X}X+2\frac{\dot Y}Y\right) \dot 
\phi -\frac{D^2e^{2\phi }}{X^2}=0.
\end{equation}
Using the time coordinate (II.22) equation (IV.16) becomes 
\begin{equation}
\phi _{,\tau \tau }+D^2Y^4=0.
\end{equation}
The equations (IV.11)-(IV.13) now become 
\begin{eqnarray}
\left( \frac{X_{,\tau }}X\right) _{,\tau }+2\left( \frac{Y_{,\tau }}Y\right)
_{,\tau }-2\frac{Y_{,\tau }}Y\left( \frac{Y_{,\tau }}Y+2\frac{X_{,\tau }}X%
\right) +2\phi _{,\tau }\left( \frac{X_{,\tau }}X+2\frac{Y_{,\tau }}Y\right)
-\frac 32\phi _{,\tau \tau }-\phi _{,\tau }^2 &=&0, \\
\left( \frac{X_{,\tau }}X\right) _{,\tau } &=&0, \\
\left( \frac{Y_{,\tau }}Y\right) _{,\tau }-\frac 12\phi _{,\tau \tau
}+kX^2Y^2e^{-2\phi } &=&0,
\end{eqnarray}
and (IV.19) and (IV.20) can be rewritten as 
\begin{eqnarray}
\left( \ln {X}\right) _{,\tau \tau } &=&0, \\
\left( \ln {Y^2e^{-\phi }}\right) _{,\tau \tau }+2kX^2Y^2e^{-2\phi } &=&0.
\end{eqnarray}
The solution of (IV.21) is 
\begin{equation}
X(\tau )=\exp {(r\tau +s)},
\end{equation}
with $r$ and $s\ $constants. For $k=0$ it is also possible to solve (IV.22)
to obtain 
\begin{equation}
Y(\tau )=\exp {\{\frac 12\left[ \phi (\tau )+m\tau +n\right] \}},
\end{equation}
with $m$ and $n$ constants. Using (IV.23)-(IV.24) we may solve
(IV.17)-(IV.20) for $Y$ and $\phi $ to give 
\begin{eqnarray}
Y(\tau ) &=&2^{\frac 14}\sqrt{\frac bD}\left[ \cosh {b\sqrt{2}\tau }\right]
^{-\frac 12}, \\
\phi (\tau ) &=&\phi _0-m\tau -\ln {\cosh {b\sqrt{2}\tau }},
\end{eqnarray}
with $\phi _0$ constant, and 
\begin{equation}
b^2=m(m+2r).
\end{equation}
Using (II.22) we have 
\begin{equation}
\tau (t)=A+\ln {t^{\frac 1{r+m}}},
\end{equation}
where 
\begin{equation}
A=\frac 1{r+m}\left[ \ln {\frac{D(r+m)}{b\sqrt{2}}}-s-\phi _0\right] .
\end{equation}
Finally, the solution of (IV.17)-(IV.20) in terms of the cosmic time $t$ in
the string frame is given by 
\begin{eqnarray}
X(t) &=&t^{\frac r{r+m}}, \\
Y(t) &=&2^{\frac 14}\sqrt{\frac bD}\left( t_0t^w+\frac 1{t_0}t^{-w}\right)
^{-\frac 12}, \\
e^{\phi (t)} &=&e^{-ma}t^{-\frac m{r+m}}\left( t_0t^w+\frac 1{t_0}%
t^{-w}\right) ^{-1},
\end{eqnarray}
where $t_0=\exp {b\sqrt{2}A}$ and $w=b\sqrt{2}/(r+m)$ are constants.

This is an axisymmetric (LRS) subcase of the Bianchi type I axion-dilaton
solution first given by Copeland et al. \cite{ed} (compare their
Eqs.(2.45)-(2.48) for $a_1=a_2$). The axisymmetric limit in \cite{ed} is
given by taking their constant $r$ to be equal to zero in their Eqs.(2.46)
and (2.47) or $C_1=C_2$ in their Eq.(2.36).

Some special solutions for the $k\neq 0$ cases can be given by solving
(IV.17) for $Y$; that is, by taking 
\begin{equation}
Y^2=\frac 1D\left( -\phi _{,\tau \tau }\right) ^{\frac 12},
\end{equation}
and substituting into (IV.20), to obtain 
\begin{equation}
\left[ \ln {\{\left( -\phi _{,\tau \tau }\right) ^{\frac 12}D^{-1}e^{-\phi }}%
\}\right] _{,\tau \tau }+2kD^{-1}\left( -\phi _{,\tau \tau }\right) ^{\frac 1%
2}e^{-2\phi (\tau )+2r\tau +2s}=0.
\end{equation}
We now seek solutions of the form 
\begin{equation}
\phi (\tau )=\alpha ^2\ln {\tau }+\beta \tau +\epsilon ,
\end{equation}
with $\alpha ^2,\beta ,$ and $\epsilon $ constants. From the field equations
(IV.17)-(IV.20) we have 
\begin{eqnarray}
X(\tau ) &=&\exp {(r\tau +s)}, \\
\phi (\tau ) &=&\frac 12\ln {\tau }+r\tau +\epsilon , \\
Y(\tau ) &=&\pm 2^{-\frac 14}D^{-\frac 12}\tau ^{-\frac 12},
\end{eqnarray}
with a constraint 
\begin{equation}
e^{2(\epsilon -s)}=\pm \frac{4k}{3D\sqrt{2}},
\end{equation}
which, because $D>0$, means we take the plus sign for $k=+1$ and the minus
sign for $k=-1$ universes. Using (II.22) and (IV.36)-(IV.38), we write 
\begin{equation}
t(\tau )=\mp D^{-1}\sqrt{2}e^{s-\epsilon }\tau ^{-\frac 12}.
\end{equation}
After deparametrisation, our solutions (IV.36)-(IV.38) give 
\begin{eqnarray}
X(t) &\propto &\exp {\left( \frac r{t^2}+s\right) }, \\
Y(t) &\propto &t, \\
\phi (t) &\propto &\ln {t}+\frac{const.}{t^2}.
\end{eqnarray}

\section{Discussion}

In this paper we have considered the low-energy-effective-action string
equations for a Kantowski-Sachs background spacetime. We have included the
full bosonic spectrum of fields, with the graviton $g_{\mu \nu }$, dilaton $%
\phi ,$ and the axion $H$. We consider the two forms of ansatz for the
axion.In terms of the pseudoscalar axion field $h,$they correspond to it
depending on either the time or space coordinates alone.

For the time-dependent case, $h=h(t),$ we found an exact parametric solution
of the field equations given in Section II. In such a case the axion field
behaves effectively as a stiff perfect-fluid (pressure = density)
distributed homogeneously over space. These solutions were also discussed in
a different context and with  more complicated parametrization in \cite{dave}%
.

We also find that, for vanishing axion $H=0$, there is a deparametrised
exact solution for $\dot \phi =\dot X/X$. We discuss this solution in
Section III. It appears to be the most interesting Kantowski-Sachs solution
in which to study the duality problem, which we shall address to a separate
paper.

For the spatially-dependent case, $h=h(r,\theta ,\psi ),$ we find that there
is only one possible form for the torsion field in spatially homogeneous
closed universes of Kantowski-Sachs type. Its 3-form strength can have just
one nonzero component, $H_{023},$ which distributes the field along the two
spatial directions on the 2-sphere $S^2$. This component depends both on
time and space and leads to space and time dependence of the only nonzero
component of the tensor potential, $B_{23}=B_{23}(t,\theta )$. This is
expected because we are working in coordinate frames rather than in
orthonormal frames of refs.\cite{bk,bar1}. In effect, there is an
anisotropic stress in the universe. For such a dilaton-axion anisotropic
cosmology we write down the field equations and find some new solutions. In
the zero-curvature case we recover the axisymmetric Bianchi I (LRS)
solutions given in \cite{ed}. These results provide, in particular, a new
type of closed universe in string cosmology.

\section{Acknowledgments}

The authors would like to thank Kerstin Kunze, Amithaba Lahiri and David
Wands for useful discussions. MPD thanks the Royal Society for support while
at the University of Sussex. MPD was also supported by the Polish Research
Committee (KBN) grant No 2 PO3B 196 10. JDB is supported by the PPARC.

\appendix

\section{Three-index axion field and pseudoscalar axion field notations}

\setcounter{equation}{0}

In this Appendix we connect our notation for the three-index axion field $H$
to that of \cite{ed} which uses the pseudoscalar axion field $h$. Following 
\cite{ed}, we define 
\begin{equation}
H^{\mu \nu \alpha }=e^\phi \epsilon ^{\mu \nu \alpha \beta }h_{,\beta }.
\end{equation}
and 
\begin{equation}
\epsilon ^{\mu \nu \alpha \beta }=\frac{4!}{\sqrt{-g}}\delta _{[0}^\mu
\delta _1^\nu \delta _2^\alpha \delta _{3]}^\beta .
\end{equation}
The equation of motion for the $h$-field is obtained via the integrability
conditions as 
\begin{equation}
\nabla ^\mu \nabla _\mu h+\nabla ^\mu \phi \nabla _\mu h=0.
\end{equation}
Notice that for the antisymmetric tensor potential, $B_{\mu \nu }=B_{\mu \nu
}(x),$ the $h$ field can only depend on time, and equation (A.3) reads 
\begin{equation}
\ddot h+\left( \frac{\dot X}X+2\frac{\dot Y}Y\right) \dot h+\dot \phi \dot h%
=0,
\end{equation}
which integrates to give 
\begin{equation}
\dot h=-A\frac{e^{-\phi }}{XY^2}
\end{equation}
So, from (A.1), we have 
\begin{equation}
H^{123}=-\frac A{X^2Y^4S(\theta )},
\end{equation}
or 
\begin{equation}
H_{123}=AS(\theta ),
\end{equation}
as required by (II.5). With the $H$ field chosen as above, the equation of
motion (II.3) is satisfied. There is also a trivial solution of (A.4), $\dot 
h=0$, but it corresponds to a constant torsion field. We assume that the $h$
field cannot depend on time at all, and the equation of motion (A.3) reads 
\begin{equation}
\partial ^\mu \partial _\mu h+\Gamma _{\rho \mu }^\mu \partial ^\rho h=0.
\end{equation}
For the Kantowski-Sachs metric, (II.4), this reads 
\begin{equation}
\frac 1{X^2}\partial _1^2h+\frac 1{Y^2}\partial _2^2h+\frac 1{Y^2S^2(\theta )%
}\partial _3^2h+\frac 1{Y^2}C(\theta )\partial _2h=0,
\end{equation}
where 
\begin{equation}
C(\theta )=\left\{ 
\begin{array}{l}
\cot {\theta }\hspace{0.5cm}{\rm for}\hspace{0.3cm}k=+1, \\ 
0\hspace{0.5cm}{\rm for}\hspace{0.3cm}k=0, \\ 
\coth {\theta }\hspace{0.5cm}{\rm for}\hspace{0.3cm}k=-1,
\end{array}
\right. \ 
\end{equation}
However, from the vanishing of the off-diagonal components of the
energy-momentum tensor which is the necessary condition to match homogeneous
Kantowski-Sachs geometry with the axion field we have 
\begin{equation}
e^\phi g^{jj}\partial _ih\partial _jh=0.
\end{equation}
This means that only one of the three $\partial _ih$ may be non-zero. The
solutions of the equation of motion (A.9) which satisfy the condition (A.11)
are as follows 
\begin{eqnarray}
\partial _1h &=&D=const.,\partial _2h=\partial _3h=0, \\
\partial _3h &=&B=const.,\partial _1h=\partial _2h=0, \\
\partial _2h &=&\frac C{S(\theta )},\partial _1h=\partial _3h=0,
\end{eqnarray}
with $C\ $constant and $S(\theta )$ given by (II.5). From (A.1), we see that
these correspond to the components of the axion field $H$ given by (IV.9)
and (IV.3) for (A.12) and (A.13) respectively, while the situation for
(A.14) is more complicated. This is because from the definition (A.1) we
have for (A.14) 
\begin{equation}
H^{013}=\frac{Ce^\phi }{XY^2S^{(}\theta )},
\end{equation}
which differs in denominator from (IV.6). With this choice, the axion
equation of motion (II.3) reads 
\begin{equation}
\partial _0\left( \frac C{S(\theta )}\right) =0,
\end{equation}
and is also fulfilled. However, we find 
\begin{equation}
H^2=H_{013}H^{013}=\frac{e^{2\phi }}{Y^2S^2(\theta )}.
\end{equation}
This means that $H$ depends on $\theta $ and the ansatz (A.14) is
inappropraite for a spatially homogeneous cosmology - a result which has
been already discussed in Section IV. The only possible solution of (A.9)
that remains is $\partial _2h=C=0$, which is a trivial solution with
constant axion. This means that there is only one consistent choice of the
components of the axion field $H,$ and this is 
\begin{equation}
H_{023}=0.
\end{equation}
However, by the definition of axion field strength, this implies that 
\begin{equation}
H_{023}=\partial _0B_{23}\neq 0,
\end{equation}
and then, bearing in mind (IV.9), it follows that the antisymmetric tensor
field potential $B_{\mu \nu }$ must depend both on time and space
coordinate, i.e., 
\begin{equation}
B_{\mu \nu }=B_{\mu \nu }(t,\theta )
\end{equation}
in a Kantowski-Sachs model. This is acceptable because the only quantity
which should be homogeneous (depending only on time) for homogeneous
geometry is the energy-momentum tensor expressible in terms of axion $H$ or
the pseudoscalar axion field $h$ (cf. Eqs. (II.1), (II.2) and \cite{ed})
which is the case for our choice (A.18). This difference appears here
because we have worked in coordinate frames rather than in orthonormal
frames of refs.\cite{bk,bar1}. Copeland et al \cite{ed} used coordinate
frames in their calculations of Bianchi I model, but it did not really
matter because the basis forms in  type I are just $\sigma ^1=dx^1,\sigma
^2=dx^2,\sigma ^3=dx^3,$ and do not involve any spatial dependence. One
could of course elaborate the problem in orthonormal frames \cite{ryan}
coming to the same conclusion as in the coordinate frames.

\newpage

\begin{center}
{\bf Figure Captions}
\end{center}

Fig.1.\\The plots of the scale factor $X$ (Eq. II.35) in terms of the parametric time
$\tau$ defined by II.22. The plots do not depend on the spatial curvature index
$k$. Different shapes of the plots depend on the values of the constants
$\alpha = \pm 1$ and $p = \pm 1$.

Fig.2.\\The plots of the scale factor $Y$ (Eq. II.37) in terms of the
parametric time $\tau$ for spatial curvature index $k = +1$. the plots depend
just on the two constants $\alpha = \pm 5$ and $p = \pm 1$ since the third
constant $\beta$ is constrained by II.41.

Fig.3.\\The plots of the dilaton $\phi$ (Eq. II.24) in terms of the parametric
time $\tau$ for different values of $\alpha = \pm 1$.

Fig.4.\\The plot of the scale factor $X(t)$ for an exact Kantowski-Sachs
low-energy-effective-action model III.9. We take $\alpha = \pm 1$ and the
ranges of $t$ are given by III.12-III.13.

Fig.5.\\The plot of the scale factor $Y(t)$ for an exact Kantowski-Sachs
low-energy-effective-action model III.10 ($\alpha = \pm 1$). The qualitative 
volume evolution (III.14) has behaviour of the same form.

Fig.6.\\The plot of the dilaton field $\phi(t)$ for an exact Kantowski-Sachs
low-energy-effective-action model III.11 ($\alpha = \pm 1$).


\begin{references}
\bibitem{frad}  E.S. Fradkin and A.A. Tseytlin {\it Nucl. Phys.} {\bf B261},
1, (1985).

\bibitem{call}  C.G. Callan, D. Friedan, E.J. Martinec and M.J. Perry, {\it %
Nucl. Phys.} {\bf B262}, 593, (1985).

\bibitem{mel}  M. Mueller, {\it Nucl. Phys.} {\bf 37}, 37, (1990).

\bibitem{ven}  A. Veneziano, {\it Phys. Lett.} {\bf B265}, 287, (1991).

\bibitem{gave}  M. Gasperini and A. Veneziano, {\it Mod. Phys. Lett.} {\bf 39%
}, 3701, (1993).

\bibitem{gave1}  M. Gasperini and A. Veneziano, {\it Astro. Phys.} {\bf 1},
317 (1993).

\bibitem{gave2}  M. Gasperini and A. Veneziano, {\it Phys. Rev. } {\bf D50},
2519, (1995).

\bibitem{ed1}  E.J. Copeland, A. Lahiri and D. Wands, {\it Phys. Rev.} {\bf %
D50}, 4868, (1994).

\bibitem{ed}  E.J. Copeland, A. Lahiri and D. Wands, {\it Phys. Rev.} {\bf %
D51}, 1569, (1995).

\bibitem{bk}  N.A. Batakis and A. Kehagias, {\it Nucl. Phys.} {\bf 449},
248, (1995).

\bibitem{gasp}  M. Gasperini and R. Ricci, {\it Class. Quantum Grav.} {\bf 12%
}, 677, (1995).

\bibitem{bar1}  J.D. Barrow and K.E. Kunze, Preprint hep-th/9608045.

\bibitem{saar}  A. Saaryan, {\it Astrophysics} {\bf 38}, 164, (1995).

\bibitem{gal}  D.V. Gal'tsov and D.V. Kechkin, {\it Phys. Rev.} {\bf D50},
        7394, (1994), {\it Phys. Lett.} {\bf B361}, 52, (1995), Preprint
        gr-qc/9608023. D.V. Gal'tsov, {\it Phys. Rev. Lett.} {\bf 74}, 2863,
        (1995).

\bibitem{kom}  A.S. Kompanyeets and A.S. Chernov, {\it Sov. Phys. JETP} {\bf %
20}, 1303, (1964).

\bibitem{KS}  R. Kantowski and R. K. Sachs, {\it J. Math. Phys.} {\bf 7}
443, (1966).

\bibitem{col}  C.B. Collins, {\it J. Math. Phys.} {\bf 18}, 2116, (1977).

\bibitem{ryan}  M.P. Ryan and L.C. Shepley, {\it Homogeneous Relativistic
Cosmologies}, Princeton U.P., Princeton, 1975.

\bibitem{dave}  J. Mim\'oso and D. Wands, {\it Phys. Rev.}D{\it \ }{\bf 51},
477 (1995).
\end{references}
\end{document}